\batchmode
\makeatletter
\def\input@path{{C:/Arbeiten/}}
\makeatother
\documentclass[english]{article}
\usepackage[T1]{fontenc}
\usepackage[latin9]{inputenc}
\usepackage{amsmath}
\usepackage{amssymb}
\usepackage{stackrel}
\usepackage{babel}
\begin{document}
\title{On the Form of Solutions of Fuchsian differential Equations with $n$
regular singular Points}
\author{Albert Huber\thanks{ahuber@tph.itp.tuwien.ac.at}}
\date{{\footnotesize{}Institut für theoretische Physik, Technische Universität
Wien, Wiedner Hauptstr. 8-10, A-1040 Wien, Austria}}
\maketitle
\begin{abstract}
The form of the coefficients of power series expressions corresponding
to solutions of Fuchsian differential equations (or their associated
degenerated confluent forms) with $n$ regular singular points is
determined by solving the corresponding $n$-term recurrence relations
in full generality. Some important special cases are discussed in
which the solutions coincide with special functions of mathematical
physics.

\medskip{}
\end{abstract}
\textit{\footnotesize{}Key words: Fuchsian differential equations,
regular singular points, recurrence relations}{\footnotesize\par}

\section{Introduction: The Theory of Fuchsian Differential Equations }

In the theory of ordinary second-order linear differential equations
with variable coefficients, which are of great interest both for mathematics
and theoretical physics, it may happen that the corresponding coefficients
are not globally well-behaved analytic functions but instead have
singularities.

In this rather common case, the behavior of solutions is usually studied
in the immediate vicinity of those (typically isolated) singular points,
where at least one of the coefficients of the equation diverges and
it is therefore to be expected - based on the fact that the corresponding
singular points of said solutions lie amongst those of their associated
singular coefficients - that at least one of the linearly independent
solutions strives toward infinity as well. The manner in which this
occurs in detail depends very much on the nature of the singular point
examined, that is, in particular, on whether the point in question
is a so-called regular or irregular singular point. 

A particularly important class of ordinary differential equations
of the aforementioned type, which deals exclusively with the much
simpler case of regular singular points, is the famous Fuchsian class
of second-order linear ordinary differential equations; a class being
specified by the property that its coefficients are (typically) complex-valued
functions that have poles of at most first and second order. Famous
representatives of this class are Gauß' hypergeometric differential
equation, Heun's equation, the Lamé equation and the generalized Lamé
equation \cite{bateman1953higher,slavjanov2000special}. 

To explain the main characteristics of this class, one may first look
at the fact that any singular ordinary homogeneous second-order linear
differential equation with variable coefficients can be be written
in the form

\begin{equation}
f''+pf'+qf=0,
\end{equation}
where the validity of the initial conditions $f\vert_{\xi=\xi_{i}}=w_{0},\;f'\vert_{\xi=\xi_{i}}=w_{1}$
is assumed in relation to a fixed non-singular point $\xi_{i}$, the
prime denotes differentiation with respect to the (typically) complex
variable $\xi$ and the coefficients $p=p(\xi)$ and $q=q(\xi)$ are
(typically) complex-valued functions with altogether $n$ different
isolated singularities $\xi_{n}$. The functions $p(\xi)$ and $q(\xi)$
are therefore regular everywhere except in the immediate vicinity
of each singular point. 

The main problem in this regard is to find a solution around a fixed
singular point$\xi_{i}$. Using the method of successive approximation
for this purpose, it can be shown that the system of first order differential
relations $f'=u$ and $u'=-pu+qf$ defined by $(1)$ has a unique
regular solution everywhere except in the vicinity of those isolated
points where $p(\xi)$ and $q(\xi)$ become singular \cite{smirnov1964lehrgang}.
Consequently, serious problems occur only close to the singular points
$\xi_{1}$, $\xi_{2}$, ..., $\xi_{n}$, so that it can be concluded
that the real difficulty is to find a solution around these points.

Accordingly, if, for the purpose of solving $(1)$ around a fixed
singlular point, it is assumed that $\xi_{i}$ that the coefficients
$p(\xi)$ and $q(\xi)$ can be expanded in Laurent series of the form
$p(\xi)=\overset{\infty}{\underset{k=-\infty}{\sum}}p_{k}(\xi-\xi_{i})^{k}$
and $q(\xi)=\overset{\infty}{\underset{k=-\infty}{\sum}}q_{k}(\xi-\xi_{i})^{k}$
in an annulus of radius $0<\vert\xi-\xi_{i}\vert<R$, it follows from
general considerations on the theory of differential equations with
variable coefficients \cite{dirschmid1976einf,smirnov1964lehrgang}
that the solutions of $(1)$ must have the form
\begin{equation}
f_{1}(\xi)=\overset{\infty}{\underset{k=-\infty}{\sum}}u_{k}(\xi-\xi_{i})^{k+\sigma_{1}},\;f_{2}(\xi)=\overset{\infty}{\underset{k=-\infty}{\sum}}v_{k}(\xi-\xi_{i})^{k+\sigma_{2}}
\end{equation}
in the case that $\sigma_{1}\neq\sigma_{2}$ or
\begin{equation}
f_{1}(\xi)=\overset{\infty}{\underset{k=-\infty}{\sum}}u_{k}(\xi-\xi_{i})^{k+\sigma_{1}},\;f_{2}(\xi)=\overset{\infty}{\underset{k=-\infty}{\sum}}v_{k}(\xi-\xi_{i})^{k+\sigma_{2}}+aw_{1}\ln(\xi-\xi_{i})
\end{equation}
in the case that $\sigma_{1}-\sigma_{2}$ is an integer.

The necessary conditions for a point $\xi_{i}$ to be a regular singular
point are that $p(\xi)$ has a pole of at most first order and $q(\xi)$
one of at most second order, so that the expressions $\underset{\xi\rightarrow\xi_{i}}{\lim}(\xi-\xi_{i})p$
and $\underset{\xi\rightarrow\xi_{i}}{\lim}(\xi-\xi_{i})^{2}q$ remain
finite. As a result, differential equation $(1)$ can be re-written
in the form
\begin{equation}
f''+\frac{p_{1}}{\xi-\xi_{i}}f'+\frac{q_{1}}{(\xi-\xi_{i})^{2}}f=0.
\end{equation}
A prerequisite for this to be the case is that the complex functions
$p_{1}(\xi)=\overset{\infty}{\underset{j=0}{\sum}}p_{j}(\xi-\xi_{i})^{j}$
and $q(\xi)=\overset{\infty}{\underset{j=0}{\sum}}q_{j}(\xi-\xi_{i})^{j}$
are holomorphic in an annulus $K$ of radius $R$, i.e. in the local
domain $\vert\xi-\xi_{i}\vert<R$ (including the center of the disc),
for arbitrary linear coefficients $p_{j}$ and $q_{j}$. 

The relation above can also be written down in the form
\begin{equation}
f''+\overset{\infty}{\underset{i=1}{\sum}}\frac{\gamma_{i}}{\xi-\xi_{i}}f'+\frac{V}{\overset{\infty}{\underset{i=1}{\prod}}(\xi-\xi_{i})}f=0,
\end{equation}
where the $\gamma_{i}$ are constant coefficients and $V=V(\xi)$
is a polynomial of degree $n-2$, usually called the Van Vleck polynomial. 

One of the main differences to differential equations with irregular
singular points, which show a stronger singular behavior, is that
the coefficients of a differential equation with regular singular
points can be expanded in the vicinity of a singular point $\xi_{i}$
(for fixed $i$) in a Laurent series with a finite instead of an infinite
number of negative exponents. Accordingly, one of the two linearly
independent solutions must be of the form
\begin{align}
f_{1}(\xi) & =\overset{\infty}{\underset{k=-m}{\sum}}u_{k}(\xi-\xi_{i})^{k+\sigma_{1}}=(\xi-\xi_{i})^{\sigma_{1}-m}(u_{-m}+u_{-m+1}(\xi-\xi_{i})+...)=:\nonumber \\
 & =:\overset{\infty}{\underset{k=0}{\sum}}w_{k}(\xi-\xi_{i})^{k+\sigma_{1}-m}
\end{align}
in full accordance with Fuchs' theorem \cite{dirschmid1976einf,smirnov1964lehrgang}.
Hence, it becomes clear that one of the solutions of $(4)$ (resp.
$(5)$) around a given singular point $\xi=\xi_{i}$ for fixed $i$
can be determined by using Frobenius' method, i.e. by making a generalized
power series ansatz of the form
\begin{equation}
f_{1}(\xi)=\overset{\infty}{\underset{k=0}{\sum}}w_{k}(\xi-\xi_{i})^{k+\rho_{1}},
\end{equation}
where the so-called critical exponent $\rho_{1}$ must be determined
by solving the indicial equation arising in the course of the calculation. 

In the case of the second solution, on the other hand, due to the
fact that $(5)$ is a linear differential relation, the method of
variation of parameters can be used to obtain another solution of
the form $f_{2}=f_{1}\overset{\xi}{\underset{\xi_{0}}{\int}}\frac{C}{f_{1}^{2}}e^{-\overset{\xi'}{\underset{\xi'_{0}}{\int}}p(\xi'')d\xi''}d\xi'$.
This again gives an expression of the form $(7)$, but for a different
critical exponent $\rho_{2}$. However, based on the fact that the
integrant of this second solution can also be expanded in a generalized
Frobenius series, it is not hard to realize - in the event that $\rho_{2}=\rho_{1}+m$,
where $m$ is some positive integer - that said solution is of the
form
\begin{equation}
f_{2}(\xi)=A\overset{\infty}{\underset{k=0}{\sum}}w_{k}(\xi-\xi_{j})^{k+\rho_{1}}\ln(\xi-\xi_{j})+\overset{\infty}{\underset{k=0}{\sum}}v_{k}(\xi-\xi_{j})^{k+\rho_{2}},
\end{equation}
where $A$ is an arbitrary constant. Of course, the corresponding
power series have positive radii of convergence. A rescaling of the
solutions $f_{1}(\xi)$ and $f_{2}(\xi)$ of $(3)$ by a factor of
$(\xi-\xi_{j})^{\alpha}$, where $\alpha$ is some positive integer,
yields again an equation of type $(5)$.

Another related way of approaching the problem solving differential
equation $(5)$ is to bring the equation into the form
\begin{equation}
\overset{n}{\underset{i=1}{\prod}}(\xi-\xi_{i})f''+\overset{n}{\underset{i=1}{\prod}}(\xi-\xi_{i})(\overset{n}{\underset{i=1}{\sum}}\frac{\gamma_{i}}{\xi-\xi_{i}})f'+Vf=0
\end{equation}
and to seek polynomial solutions of the resulting expression. These
solutions, if they exist, are then given by what are called Heine--Stieltjes
polynomials (sometimes also called Stieltjes polynomials) \cite{heine1878handbuch,marden1931stieltjes,stieltjes1885certains,volkmer1999expansions},
which form the basis for the construction of ellipsoidal harmonics
and their generalizations \cite{volkmer2006generalized}. In the case
that no such polynomial solutions of $(9)$ exist, the solution will
be of the form 

\begin{equation}
f_{1}(\xi)=\overset{\infty}{\underset{k=0}{\sum}}w_{k}\xi{}^{k+\rho_{1}}.
\end{equation}
In order to obtain the second possible solution with the aid of the
already determined one, one may then either use ansatz $(8)$ or possibly
first seek another solution by applying the method of variation of
parameters in a slightly different way, namely by trying to find a
solution $\Delta=\Delta(\xi)$ of $(9)$ for $V=0$ and then make
an ansatz of the form 
\begin{equation}
f_{2}=f_{1}\Delta+G,
\end{equation}
where $G=G(\xi)$ can be assumed to be a generalized power series
of the form $G(\xi)=\overset{\infty}{\underset{k=0}{\sum}}v_{k}\xi{}^{k+\rho_{2}}$.
Provided that the definition $\mho\equiv\Delta'$ is used in the present
context, $\Delta(\xi)$ is then obtained by solving the first order
relation
\begin{equation}
\mho'+\overset{n}{\underset{j=1}{\sum}}\frac{\gamma_{j}}{\xi-\xi_{j}}\mho=0,
\end{equation}
which follows directly from $(9)$ for the special case $V(\xi)=0$;
leading to the result $\Delta=C\int\mho d\xi$, where $\mho=\mho(\xi)=e^{-\overset{n}{\underset{j=1}{\sum}}\gamma_{j}\ln(\xi-\xi_{j})}$
and $\Delta=\Delta(\xi)=\frac{df_{1}}{d\xi}f_{2}-f_{1}\frac{df_{2}}{d\xi}$
is, of course, the Wronski determinant. 

Both methods are equivalent in that they lead to exactly the same
results, from which it can be concluded that the respective pairs
$(7)$ and $(8)$ and $(10)$ and $(11)$ can be superimposed in such
a way that they represent one and the same solution of $(5);$ just
written down in different ways. The concrete choice of one of these
methods for solving Fuchsian differential equations is therefore purely
a matter of taste. 

Anyway, the question of the convergence of the corresponding power
series expressions still remains to be clarified. For the sake of
simplicity, $\xi_{i}=0$ shall be assumed at this point. Given this
case, equation $(4)$ can be re-written in the form 
\begin{equation}
\xi^{2}f''+\xi\overset{n}{\underset{j=0}{\sum}}p_{j}\xi{}^{j}f'+\overset{n}{\underset{j=0}{\sum}}q_{j}\xi{}^{j}f=0.
\end{equation}
Using ansatz $(7),$ which in the given case coincides with $(10)$,
the system of equations 
\begin{equation}
\begin{cases}
 & w_{0}f_{0}(\rho_{1})=0;\\
 & w_{1}f_{0}(\rho_{1}+1)+w_{0}f_{1}(\rho_{1})=0;\\
 & w_{2}f_{0}(\rho_{1}+2)+w_{1}f_{1}(\rho_{1}+1)+w_{0}f_{2}(\rho_{1})=0;\\
 & ................................................................................\\
 & w_{n}f_{0}(\rho_{1}+n)+w_{n-1}f_{1}(\rho_{1}+n-1)+...+w_{0}f_{n}(\rho_{1})=0;\\
 & ......................................................................................................,
\end{cases}
\end{equation}
is obtained for the coefficients $w_{k}$, where the abbreviations
$f_{0}(x)=x(x-1)+p_{0}x+q_{0}$ and $f_{k}(x)=p_{k}x+q_{k}$ were
introduced for the sake of brevity. If $R$ is the radius of convergence
of the coefficients occurring in $(13)$ and $R_{1}<R$, one can use
the estimates $\vert p_{k}\vert<\frac{m_{1}}{R_{1}^{k}}$ and $\vert q_{k}\vert<\frac{m_{2}}{R_{1}^{k}}$
to conclude that $\vert p_{k}\vert+\vert q_{k}\vert<\frac{m_{1}+m_{2}}{R_{1}^{k}}<\frac{M}{R_{1}^{k}}$
applies in the present context, where $M$ is assumed to be some sufficiently
large number. In addition, one can use the fact that $\underset{n\rightarrow\infty}{\lim}\frac{\vert\rho_{1}\vert+n}{f_{0}(\rho_{1}+n)}\rightarrow0$
implies that there is always a positive integer $n\geq N$ such that
$\vert f_{0}(\rho_{1}+n)\vert>\vert\rho_{1}\vert+n$. Considering
then the fact that $\vert f_{k}(\rho_{1}+n-k)\vert\leq\vert q_{k}\vert+(\rho_{1}+n-k)\vert p_{k}\vert<\vert q_{k}\vert+(\vert\rho_{1}\vert+n)\vert p_{k}\vert<(\vert\rho_{1}\vert+n)(\vert q_{k}\vert+\vert p_{k}\vert)$,
it can be concluded that $\frac{\vert f_{k}(\rho_{1}+n-k)\vert}{\vert f_{0}(\rho_{1}+n)\vert}<\frac{M}{R_{1}^{k}}$
must hold. With these result at hand, $(14)$ can be used to show
that $\vert w_{n}\vert\leq\frac{\vert f_{1}(\rho_{1}+n-1)\vert}{\vert f_{0}(\rho_{1}+n)\vert}\vert w_{n-1}\vert+\frac{\vert f_{2}(\rho_{1}+n-2)\vert}{\vert f_{0}(\rho_{1}+n)\vert}\vert w_{n-2}\vert+...+\frac{\vert f_{n}(\rho_{1})\vert}{\vert f_{0}(\rho_{1}+n)\vert}\vert w_{0}\vert$
and thus $\vert w_{n}\vert\leq\frac{M}{R_{1}}\vert w_{n-1}\vert+\frac{M}{R_{1}^{2}}\vert w_{n-2}\vert+...+\frac{M}{R_{1}^{n}}\vert w_{0}\vert$.
For the first $N$ coefficients, it is always possible find a sufficiently
large number $P>1+M$ such that one can estimate $\vert w_{k}\vert\leq\frac{P^{k}}{R_{1}^{k}}$
for $k=1,2,...,N-1$. For fixed $n\geq N$, using the fact that $P^{n}(P-M-1)+M>0$,
one then obtains $\vert w_{n}\vert\leq\frac{M}{R_{1}^{n}}(P^{n-1}+P^{n-2}+...+1)=\frac{P^{n}-1}{P-1}\frac{M}{R_{1}^{n}}\leq\frac{P^{n}}{R_{1}^{n}}$.
The estimate thus remains correct for any given value of $n$. However,
since the series $\overset{\infty}{\underset{k=0}{\sum}}\frac{P^{k}}{R_{1}^{k}}\xi^{k}$
is absolutely convergent in an annulus of radius $\vert\xi\vert<\frac{R_{1}}{P}$,
this proves that the generalized power series expression $(10)$ represents
a solution of $(4)$ in the immediate vicinity of $\xi_{0}=0$. And
since analog results can be deduced in the immediate vicinity of all
other singular points, one comes to the conclusion that $(7)$ indeed
represents a solution of the differential equations $(4)$ and $(5)$.

Having clarified this, one can take now advantage of the fact that,
by analytic continuation along any path not passing through the poles
of $p(\xi)$ and $q(\xi)$, any set of linearly independent solutions
being valid around a singular point of differential equation $(5)$
gives a new set of solutions. However, it typically occurs in this
context that the extended functions obtained from the analytic continuation
of a previously given set of solutions in the vicinity of an isolated
singular point $\xi=\xi_{i}$ are multivalued complex functions, whose
value at another point $\xi=\xi_{j}$ depends on the chosen curve
from $\xi_{i}$ to $\xi_{j}$. In particular, by choosing a path around
one of the singular points, it often happens that this point becomes
a branch point, which has the effect that a given pair $f_{1}(\xi)$
and $f_{2}(\xi)$ of solutions transitions into a new pair $\tilde{f}_{1}(\xi)$
and $\tilde{f}_{2}(\xi)$. However, by taking advantage of the fact
that pairs of solutions form a vector space, it becomes clear that
there is a monodromy transformation, which, after a certain singularity
has been circulated either clock- or counterclockwise, turns one pair
of solutions into another. In this context, the components of the
corresponding monodromy matrix are constants and its determinant must
be different from zero in order to ensure that the solutions $\tilde{f}_{1}(\xi)$
and $\tilde{f}_{2}(\xi)$ are linearly independent. The monodromy
matrices are the generators of the monodromy group, which is prescribed
by means of a finite-dimensional complex linear representation of
the so-called fundamental group, which is the first and simplest homotopy
group \cite{piccinini1992lectures,ravenel2003complex}. 

The monodromy concept is important in this context not least because
its definition reveals an important property of analytical continuations
along curves between regular singular points. This can be seen if
one moves alongside special paths around an isolated singularity,
which all have the same start and endpoints and can be continuously
deformed into one another, because in this case the analytic continuations
along different curves will yield the same results at their common
endpoint, which is subject to the famous monodromy theorem. 

Given the concrete form of the linearly independent solutions in the
vicinity of an isolated singular point, the question is how the form
of the solution looks at other singular points of the equation. 

Thus, in the case that an analytic continuation of a given pair of
solutions in the vicinity of a fixed regular singular point can be
defined, it is ensured that those pairs of solutions, which are valid
in the vicinity of all other regular singular points, can be converted
into each other by simple linear transformations. The analytic continuation
of solutions along curves, which 'connect' in this way pairs of regular
singular points with each other, then defines a Riemann surface, i.
e. a one-dimensional complex manifold, which is a connected Hausdorff
space that is endowed with an atlas of charts to the open unit disk
of the complex plane (whereas the transition maps between two overlapping
charts are required to be holomorphic).

Consequently, it is clear how to transform any set of solutions of
differential equation $(5)$, which is defined in the immediate vicinity
of a given singular point $\xi_{i}$, to one being defined in the
immediate vicinity of another singular point $\xi_{j}$. In this way,
the respective form of solutions can be easily calculated in the vicinity
of each isolated singular point, whereas it turns out the group of
symmetries is isomorphic to the Coxeter group $D_{n}$ of order $n!2^{n-1}$
\cite{yoshida1987fuchsian}. The total number of possible solutions
can therefore easily be given for any number of regular singular points,
which proves to be extremely interesting not only from a mathematical,
but also from a physical point of view, since a large number of important
equations of mathematical physics happen to be Fuchsian differential
equations with a (typically) small number of regular singular points.

Of course, all of this has been known for a long time. In point of
fact, the theory of Fuchsian differential equations has been a fixed,
fully established part of the mathematical theory of second-order
linear homogeneous differential equations with variable coefficients
since the late nineteenth, early twentieth century. Nevertheless,
there is an important point to be addressed here: What all Fuchsian
differential equations with more than three regular singular points
have in common is that their solutions, which have to be of the form
$(7)$ and $(8)$ are difficult to write down explicitly. More precisely,
it is hard to write down the form of the coefficients of the corresponding
generalized power series expression for any finite number of terms
of the associated recurrence relation. In particular, given the case
that relation $(9)$ is used as a starting point for the analysis
and ansatz $(7)$ is used for finding a solution around a given singular
point $\xi_{j}$, there occurs the problem that one (typically) has
to deal with a $n-2$-term recursion relation of the form
\begin{equation}
w_{k+n-2}=\overset{n-2}{\underset{j=1}{\sum}}m_{(j)k+n-2}w_{k+n-j-2},
\end{equation}
where the $m_{(j)k}=m_{(j)}(k)$ are (not necessarily smooth or even
regular) functions of $k$. Note that, of course, this is completely
in line with what is obtained from relation $(14)$.

Except for the almost trivial case $n=3$, it is often a delicate
problem to write down solutions of $(15)$. Nevertheless, as shall
now be discussed, this problem can be solved by considering special
distributionally-valued coefficients that can be used to determine
the exact form of solutions of $(9)$, i.e. of solutions that are
given in terms of power series expressions of the form $(10)$ and
$(11)$. The next section shall clarify this in full detail.

\section{Recurrence Relations and distributionally-valued Coefficients}

In the previous section, it has been pointed out that in order to
find solutions of the form $(7)$ of representatives of the class
of Fuchsian differential equations given by $(5)$, one typically
has to deal with the problem of solving associated recurrence relations
of the form $(15)$. The form of the solutions of this relation is
subject to the following theorem:

\textbf{Theorem 1: }\textit{Let $(10)$ be a solution of $(9)$ for
fixed critical exponent $\rho_{1}$ and $(15)$ be the associated
recurrence relation. Given the choice} $w_{1}:=m_{(1)0}w_{0}$, $w_{2}:=m_{(1)1}w_{1}+m_{(2)1}w_{0}=(m_{(1)1}m_{(1)0}+m_{(2)1})w_{0}$,
..., $w_{n-2}:=m_{(1)n-2}w_{n-3}+m_{(2)n-2}w_{n-4}+...+m_{(n-2)n-2}w_{0}$\textit{,
solutions of $(15)$ take the form} 
\begin{equation}
w_{k+n-2}=\ll m_{(1)},m_{(2)},...,m_{(n-2)}\gg_{k+n-2}w_{0},
\end{equation}
\textit{where the symbol} $\ll m_{(1)},m_{(2)},...m_{(n-2)}\gg_{k+n-2}$
\textit{can be written in terms of a multi-linear form of the type}
\begin{equation}
\ll m_{(1)},m_{(2)},...,m_{(n-2)}\gg_{k+n-2}\::=W_{a_{k+n-2}...a_{0}}X_{k+n-2}^{a_{k+n-2}}...X_{0}^{a_{0}}.
\end{equation}
In this context, each $a_{j}$ runs from zero to $k+n-2$ and the
corresponding objects $X_{j}^{a_{j}}=X^{a_{j}}(j)$ have the non-zero
components $X_{j}^{0}=\theta(j),$ $X_{j}^{1}=m_{(1)}(j),$ $X_{j}^{2}=m_{(2)}(j)$,
... $X_{j}^{k+n-2}=m_{(k+n-2)}(j),$ where $\theta(j):=\begin{cases}
\overset{0}{\underset{1}{}} & \overset{if\:j<0}{\underset{if\:j\geq0}{}}\end{cases}$ is the Heaviside step function. The object $W_{a_{k+n-2}...a_{0}}$
is defined in such a way that all its components are either zero or
one. The non-zero components are exactly those for which on the one
hand $\overset{k+n-2}{\underset{j=0}{\sum}}a_{j}=k+n-1$ applies and,
on the other hand, all indices that take the value zero occur only
as successors of those that take a value of two, all pairs of indices
that take the value zero combined occur only as successors of indices
with a value of three, all triples of indices that take the value
zero occur only as successors of indices with a value of four and
so on and so forth. This implies in particular that all $W_{0a_{k+n-3...}a_{0}}$,
$W_{00a_{k+n-4...}a_{0}}$, ... $W_{00...0a_{n-2}...a_{0}}$ are zero.
It is also assumed that all coefficients with negative values are
zero by definition. 

\textbf{Proof: }As a basis for proving that $(16)$ is a solution
of $(15)$, it must be proven by induction that

\begin{align}
 & W_{a_{k+n-2}...a_{0}}X_{k+n-2}^{a_{k+n-2}}...X_{0}^{a_{0}}\overset{!}{=}m_{(1)k+n-2}W_{a_{k+n-3}...a_{0}}X_{k+n-3}^{a_{k+n-3}}...X_{0}^{a_{0}}+\\
 & +m_{(2)k+n-2}W_{a_{k+n-4}...a_{0}}X_{k+n-4}^{a_{k+n-4}}...X_{0}^{a_{0}}+...+m_{(n-2)k+n-2}W_{a_{n-2}...a_{0}}X_{n-2}^{a_{n-2}}...X_{0}^{a_{0}}.\nonumber 
\end{align}
Considering the sequence of coefficients $w_{1}:=m_{(1)0}w_{0}$,
$w_{2}:=m_{(1)1}w_{1}+m_{(2)1}w_{0}=(m_{(1)1}m_{(1)0}+m_{(2)1})w_{0}$,
..., $w_{n-2}:=m_{(1)n-2}w_{n-3}+m_{(2)n-2}w_{n-4}+...+m_{(n-2)n-2}w_{0}$
in combination with relations $(16)$ and $(17)$, it immediately
becomes clear that $W_{a_{0}}X_{0}^{a_{0}}=W_{1}X_{0}^{1}=m_{(1)0}$,
$W_{a_{1}a_{0}}X_{1}^{a_{1}}X_{0}^{a_{0}}=m_{(1)1}m_{(1)0}+m_{(2)1}=m_{(1)1}W_{a_{0}}X_{0}^{a_{0}}+m_{(2)1}$,
$W_{a_{2}a_{1}a_{0}}X_{2}^{a_{2}}X_{1}^{a_{1}}X_{0}^{a_{0}}=m_{(1)2}(m_{(1)1}m_{(1)0}+m_{(2)0})+m_{(2)2}m_{(1)0}+m_{(3)0}=m_{(1)2}W_{a_{1}a_{0}}X_{1}^{a_{1}}X_{0}^{a_{0}}+m_{(2)2}W_{a_{0}}X_{0}^{a_{0}}+m_{(3)0}$,
... must hold. This gives the non-zero components $W_{1}$, $W_{11}$,
$W_{20},$ $W_{111}$, $W_{120}$, $W_{201}$, ....$\,$. Repeating
this procedure for continuously increasing numbers of $n$, it becomes
clear that the above sequence of coefficients can be combined to multi-linear
form of the type $\ll m_{(1)},m_{(2)},...,m_{(n-2)}\gg_{n-2}=W_{a_{n-2}...a_{0}}X_{n-2}^{a_{n-2}}...X_{0}^{a_{0}},$
provided that the condition $\overset{n-2}{\underset{j=0}{\sum}}a_{j}=n-1$
is met and all indices that take the value zero occur only as successors
of those that take a value of two, all pairs of indices that take
the value zero combined occur only as successors of indices with a
value of three, all triples of indices that take the value zero occur
only as successors of indices with a value of four and so on and so
forth.

The first non-trivial case to be considered is $k=1$. In this case,
the above relation reads

\begin{align}
W_{a_{n-1}a_{n-2}...a_{0}}X_{n-1}^{a_{n-1}}X_{n-2}^{a_{n-2}}...X_{0}^{a_{0}} & =W_{1a_{n-2}...a_{0}}X_{n-1}^{1}X_{n-2}^{a_{n-2}}...X_{0}^{a_{0}}+\nonumber \\
+W_{20a_{n-3}...a_{0}}X_{n-1}^{2}X_{n-2}^{0}X_{n-3}^{a_{n-3}}...X_{0}^{a_{0}}+ & W_{300a_{n-4}...a_{0}}X_{n-1}^{3}X_{n-2}^{0}X_{n-3}^{0}X_{n-4}^{a_{n-4}}...X_{0}^{a_{0}}+...=\nonumber \\
\overset{!}{=}m_{(1)n-1}W_{a_{n-2}...a_{0}}X_{n-2}^{a_{n-2}}...X_{0}^{a_{0}}+m_{(2)n-1} & W_{a_{n-3}...a_{0}}X_{n-3}^{a_{n-3}}...X_{0}^{a_{0}}+\nonumber \\
+m_{(3)n-1}W_{a_{n-4}...a_{0}}X_{n-4}^{a_{n-4}}...X_{0}^{a_{0}}+... & .
\end{align}
Given the fact that $W_{1a_{n-2}...a_{0}}-W_{a_{n-2}...a_{0}}=W_{20a_{n-3}...a_{0}}X_{n-2}^{0}-W_{a_{n-3}...a_{0}}=W_{300a_{n-4}...a_{0}}X_{n-2}^{0}X_{n-3}^{0}-W_{a_{n-4}...a_{0}}=....=0$
is fulfilled in the given case, since all non-zero components of both
$W_{a_{n-1}a_{n-2}...a_{0}}$ and $W_{a_{n-j-1}...a_{0}}$ have the
same value equal to one, assertion $(18)$ defines a distributional
relation which is actually fulfilled for all possible combinations
of indices $a_{n-2}...a_{0}$, $a_{n-3}...a_{0}$, ...., $a_{n-2-j}...a_{0}$
due to the fact that $X_{j}^{0}=1$ applies for all fixed non-negative
values $j$.

The next step is $k=2$, but this works completely the same if all
indices with values of $n-1$ and $n-2$ in $(19)$ are replaced by
indices with values $n$ and $n-1$ . 

Finally, the consistency of the induction step $k\rightarrow k+1$
can easily be verified by replacing again all indices with values
of $n-1$ and $n-2$ in $(19)$ by indices with values $k+n-1$ and
$k+n-2$, so that the validity of assertion $(18)$ follows as a direct
consequence. Thus, one obtains the result
\begin{align}
 & W_{a_{k+n-2}...a_{0}}X_{k+n-2}^{a_{k+n-2}}...X_{0}^{a_{0}}=\\
= & \overset{n-2}{\underset{j=1}{\sum}}m_{(j)k+n-2}W_{a_{k+n-j-2}...a_{0}}X_{k+n-j-2}^{a_{k+n-j-2}}...X_{0}^{a_{0}}.\nonumber 
\end{align}
After multiplication with $w_{0}$, it then becomes clear that - under
the given choice for the coefficients $w_{1}$, $w_{2}$,...,$w_{n-2}$
- $(12)$ actually represents a solution to $(15).$ 

$\hfill\qquad\qquad\qquad\qquad\qquad\qquad\qquad\Box$

\medskip{}
The additional solution $f_{2}=f_{2}(\xi)$ has a related, but more
complicated form. This form can be obtained by inserting ansatz $(10)$
for a different choice of critical exponent $\rho_{2}$ into $(9)$,
which then gives a second solution of precisely the same form. However,
in the case that the corresponding critical exponents differ by a
positive integer, ansatz $(11)$ can be used, which then leads to
the differential relation 
\begin{equation}
\overset{n}{\underset{j=1}{\prod}}(\xi-\xi_{j})G''+\overset{n}{\underset{j=1}{\prod}}(\xi-\xi_{j})(\overset{n}{\underset{j=1}{\sum}}\frac{\gamma_{j}}{\xi-\xi_{j}})G'+VG+2\overset{n}{\underset{j=1}{\prod}}(\xi-\xi_{j})\mho f_{1}'=0
\end{equation}
in the power series $G(\xi)=\overset{\infty}{\underset{k=0}{\sum}}v_{k}\xi{}^{k+\rho_{2}}$.
By expanding the term $\overset{n}{\underset{j=1}{\prod}}(\xi-\xi_{j})\mho f_{1}'$
in a power series as well, one then typically ends up with the recurrence
relation of the form

\begin{equation}
v_{k+n-2}=\overset{n-2}{\underset{j=1}{\sum}}m_{(j)k+n-2}v_{k+n-j-2}+\Phi_{k+n-2},
\end{equation}
where the coefficient $\Phi_{k}$ results from the expansion. The
solutions of this equation are subject to the following theorem:

\textbf{Theorem 2: }\textit{Let $(11)$ be a solution of $(9)$ for
fixed critical exponents $\rho_{1}$ and $\rho_{2}$ and $(22)$ be
the associated recurrence relation. Given the choice} $v_{1}:=m_{(1)0}v_{0}+\Phi_{0}$,
$v_{2}:=m_{(1)1}v_{1}+m_{(2)1}v_{0}+\Phi_{1}=(m_{(1)1}m_{(1)0}+m_{(2)1})v_{0}+\Phi_{1}+m_{(1)1}\Phi_{0}$,
..., $v_{n-2}:=m_{(1)n-2}v_{n-3}+m_{(2)n-2}v_{n-4}+...+m_{(n-2)n-2}v_{0}+\Phi_{n-2}+m_{(1)n-3}\Phi_{n-3}+(m_{(1)1}m_{(1)0}+m_{(2)1})\Phi_{n-4}+...$\textit{,
solutions of recurrence relation $(22)$ take the form} 
\begin{equation}
v_{k+n-2}=\;\lll\parallel m_{(1)},m_{(2)},...m_{(n-2)}\parallel\ggg_{k+n-2}
\end{equation}
\textit{where, assuming that} $\overset{k+n-j-3}{\underset{m=0}{\sum}}a_{m}=k+n-j-2$
and $\overset{k+n-2}{\underset{m=k+n-j-2}{\sum}}a_{m}=j+1$ \textit{applies
in the present context, the symbol} $\lll\parallel m_{(1)},m_{(2)},...m_{(k+n-2)}\parallel\ggg_{k+n-2}$
\textit{can be written in terms of a multi-linear form of the type}

\begin{align}
 & \lll\parallel m_{(1)},m_{(2)},...m_{(n-2)}\parallel\ggg_{k+n-2}\;=\;\ll m_{(1)},m_{(2)},...m_{(n-2)}\gg_{k+n-2}v_{0}+\nonumber \\
 & +\Phi_{k+n-2}+\overset{k+n-2}{\underset{j=0}{\sum}}W_{a_{k+n-2}...a_{k+n-j-2}}X_{k+n-2}^{a_{k+n-2}}...X_{k+n-j-2}^{a_{k+n-j-2}}\Phi_{k+n-j-3}.
\end{align}
\textbf{Proof: }The first non-trivial case to be considered is $k=1$.
In this case, it can be concluded that $(22)$ is solved if and only
if 
\begin{align}
 & \lll\parallel m_{(1)},m_{(2)},...m_{(n-2)}\parallel\ggg_{n-1}\overset{!}{=}\\
\overset{!}{=} & \overset{n-2}{\underset{j=1}{\sum}}m_{(j)n-1}\lll\parallel m_{(1)},m_{(2)},...m_{(n-2)}\parallel\ggg_{n-j-1}+\Phi_{n-1}\nonumber 
\end{align}
Using $(24)$, the left hand side of $(25)$ can be written down in
the form
\begin{align}
 & \lll\parallel m_{(1)},m_{(2)},...m_{(n-2)}\parallel\ggg_{n-1}=\ll m_{(1)},m_{(2)},...m_{(n-2)}\gg_{n-1}v_{0}+\\
 & +\Phi_{n-1}+\overset{n-2}{\underset{i=0}{\sum}}W_{a_{n-2}...a_{n-i-2}}X_{n-2}^{a_{n-2}}...X_{n-i-2}^{a_{n-i-2}}\Phi_{n-i-2},\nonumber 
\end{align}
whereas the right hand side reads

\begin{align}
 & \overset{n-2}{\underset{j=1}{\sum}}m_{(j)n-1}\ll m_{(1)},m_{(2)},...m_{(n-2)}\gg_{n-j-1}v_{0}+\Phi_{n-1}+\\
 & +\overset{n-2}{\underset{j=1}{\sum}}m_{(j)n-1}\left\{ \Phi_{n-j-1}+\overset{n-j-2}{\underset{i=0}{\sum}}W_{a_{n-j-2}...a_{n-j-i-2}}X_{n-j-2}^{a_{n-j-2}}...X_{n-j-i-2}^{a_{n-j-i-2}}\Phi_{n-j-i-2}\right\} .\nonumber 
\end{align}
Due to the validity of $(20)$, one thus is left to prove 
\begin{align}
 & \overset{n-2}{\underset{i=0}{\sum}}W_{a_{n-2}...a_{n-i-2}}X_{n-2}^{a_{n-2}}...X_{n-i-2}^{a_{n-i-2}}\Phi_{n-i-2}\overset{!}{=}\\
 & \overset{!}{=}\overset{n-2}{\underset{j=1}{\sum}}m_{(j)n-1}\left\{ \Phi_{n-j-1}+\overset{n-j-2}{\underset{i=0}{\sum}}W_{a_{n-j-2}...a_{n-j-i-2}}X_{n-j-2}^{a_{n-j-2}}...X_{n-j-i-2}^{a_{n-j-i-2}}\Phi_{n-j-i-2}\right\} .\nonumber 
\end{align}
Using here the fact that this relation can be re-written in the form
\begin{align*}
 & W_{a_{n-2}}X_{n-2}^{a_{n-2}}\Phi_{n-2}+W_{a_{n-2}a_{n-3}}X_{n-2}^{a_{n-2}}X_{n-3}^{a_{n-3}}\Phi_{n-3}+...+W_{a_{n-2}...a_{0}}X_{n-2}^{a_{n-2}}...X_{0}^{a_{0}}\Phi_{0}\overset{!}{=}\\
 & \overset{!}{=}m_{(1)n-1}(\Phi_{n-2}+W_{a_{n-3}}X_{n-3}^{a_{n-3}}\Phi_{n-3}+...+W_{a_{n-3}...a_{0}}X_{n-3}^{a_{n-3}}...X_{0}^{a_{0}}\Phi_{0})+\\
 & +m_{(2)n-1}(\Phi_{n-3}+W_{a_{n-4}}X_{n-4}^{a_{n-4}}\Phi_{n-4}+...+W_{a_{n-4}...a_{0}}X_{n-3}^{a_{n-3}}...X_{0}^{a_{0}}\Phi_{0})+\\
 & +...+\\
 & m_{(n-2)n-1}(\Phi_{1}+W_{a_{0}}X_{0}^{a_{0}}\Phi_{0})=m_{(1)n-1}\Phi_{n-2}+(m_{(1)n-1}W_{a_{n-3}}X_{n-3}^{a_{n-3}}+m_{(2)n-1})\Phi_{n-3}+...\\
 & +(m_{(1)n-1}W_{a_{n-3}...a_{0}}X_{n-3}^{a_{n-3}}...X_{0}^{a_{0}}+m_{(2)n-1}W_{a_{n-4}...a_{0}}X_{n-3}^{a_{n-3}}...X_{0}^{a_{0}}+...+m_{(n-2)n-1})\Phi_{0},
\end{align*}
one finds that it is sufficient to show that the condition
\begin{equation}
W_{a_{n-2}...a_{0}}X_{n-2}^{a_{n-2}}...X_{0}^{a_{0}}\overset{!}{=}\overset{n-2}{\underset{j=1}{\sum}}m_{(j)n-1}W_{a_{n-j-2}...a_{0}}X_{n-j-2}^{a_{n-j-2}}...X_{0}^{a_{0}}
\end{equation}
is met in the present context. However, by taking into account that
the left hand side can be written in the form
\begin{align}
 & W_{a_{n-2}...a_{0}}X_{n-2}^{a_{n-2}}...X_{0}^{a_{0}}=W_{1a_{n-3}...a_{0}}X_{n-2}^{1}X_{n-3}^{a_{n-3}}...X_{0}^{a_{0}}+\\
 & W_{20a_{n-4}...a_{0}}X_{n-2}^{2}X_{n-3}^{0}X_{n-4}^{a_{n-4}}...X_{0}^{a_{0}}+...+W_{n-200...0}X_{n-2}^{n-2}X_{n-3}^{0}X_{n-4}^{0}...X_{0}^{0},\nonumber 
\end{align}
one finds that condition $(29)$ is actually met due to the fact that
$W_{1a_{n-3}...a_{0}}-W_{a_{n-3}...a_{0}}=W_{20a_{n-4}...a_{0}}X_{n-2}^{0}-W_{a_{n-4}...a_{0}}=...=W_{n-200...0}X_{n-3}^{0}X_{n-4}^{0}...X_{0}^{0}-1=0$
holds by definition for all $n-j>0$ with $3\leq j\leq n$. 

Accordingly, since one can proceed completely the same way for any
given larger value of $k$, it can be concluded that ansatz $(23)$
can be used to solve recurrence relation $(22)$ and therefore that
$(11)$ indeed represents a solution to $(9)$ for the given choice
of coefficients. 

$\hfill\qquad\qquad\qquad\qquad\qquad\qquad\qquad\Box$

\medskip{}
As an application of the present investigation of Fuchs' mathematical
framework for solving second-order linear differential equations with
$n$ regular singular points, special types of differential equations
belonging to this class shall be considered next, which play an important
role in both mathematics and theoretical physics. Since, however,
the mathematical literature pays much attention to the discussion
of these special differential equations with a small number of regular
singular points anyway, the remaining part of this section will be
content with giving some relevant examples without discussing them
in full detail on a case-by-case basis or listing the exact structure
of the associated solutions and all their relevant properties. For
a more detailed treatment of the subject, one should therefore consult
the relevant mathematical literature, such as for instance the books
by Bateman, Slavnov and Lay or Smirnov. 

The simplest non-trivial example of a Fuchsian differential equation
is one with three regular singular points. However, as it turns out,
any equation of this type can be transformed into Gauß' hypergeometric
differential equation, which is the reason why it is not only indisputably
the most prominent, but also the only relevant representative of this
class. From a physical point of view, this is certainly because it
leads to some important special cases, including the Legendre equation,
the Jacobi equation, the Chebyshev equation and the Gegenbauer equation;
all of which possess polynomial solutions that belong to the superordinate
class of Heine-Stieltjes polynomials. By performing a linear transformation,
which serves as a basis of a limiting procedure by means of which
it can be achieved that a singularity lying at a finite position is
shifted into infinity and thus coincides with the singularity already
existing there, the said differential equation can be converted into
the so-called confluent hypergeometric differential equation, which
leads to other important special cases such as the Bessel, Hermite
and Laguerre equations. It thus becomes apparent that a large number
of the special functions relevant for mathematical physics are solutions
of the hypergeometric equation in one form or another.

The said differential equation can be written down in the form
\begin{equation}
\xi(\xi-1)f''+[(a+b+1)\xi-c]f'+abf=0,
\end{equation}
where $a$, $b$ and $c$ are arbitrary linear coefficients. It has
three singular points at $0$, $1$ and $\infty$ and critical exponents
$\rho_{1}=0$ and $\rho_{2}=1-c$. The first solution leads to the
two-term recursion relation 
\begin{equation}
w_{k+1}=m_{(1)k+1}w_{k},
\end{equation}
which contains the definition $m_{(1)k+1}:=\frac{(a+k)(b+k)}{(k+1)(c+k)}$.
This relation can trivially be solved by continuous iteration, which
gives the result
\begin{equation}
w_{k+1}=\ll m_{(1)}\gg_{k+1}w_{0},
\end{equation}
which can be re-written by using the fact that $\ll m_{(1)}\gg_{k+1}=W_{11...1}X_{k+1}^{1}X_{k}^{1}...X_{0}^{1}=\overset{k+1}{\underset{j=0}{\prod}}m_{(1)j}=\frac{(a)_{k+1}(b)_{k+1}}{(k+1)!(c)_{k+1}},$
where $(x)_{k}=\frac{\Gamma(x+k)}{\Gamma(k)}=\begin{cases}
\stackrel[1]{x(x+1)....(x+k-1)}{} & \stackrel[k=0]{k\geq1}{}\end{cases}$ is the Pochhammer symbol. The solution found is therefore given by
the hypergeometric function $f_{1}(\xi)=F(a,b;c;\xi)=\stackrel[k=0]{\infty}{\sum}\frac{(a)_{k}(b)_{k}}{(c)_{k}}\frac{\xi^{k}}{k!}$,
as it ought to be. Depending on the values for $a$, $b$ or $c$,
different cases have to be distinguished. In the case that $1-c$
is not a positive integer, the second solution reads $f_{2}(\xi)=\xi^{1-c}F(a-c+1,b-c+1,2-c;\xi)$
. In turn, given the case that it is a positive integer, one of the
solutions given above loses its meaning. In this case, as well as
in the special case in which $c=1$, one of the solutions will be
of the form $(8)$ and thus contains a logarithmic term. And while
these more complicated cases can, of course, be treated within the
formalism in the same way as all those other solutions that combine
to Kummer's collection of $24$ solutions of the hypergeometric differential
equation in the vicinity of all regular singular points $0$, $1$
and $\infty$, it seems more reasonable in the present context, in
order to demonstrate the practicality of the formalism at hand, to
pass over directly to Fuchsian differential equations with a larger
number of regular singular points. 

For this purpose, consider the case of Fuchsian differential equations
with four regular singular points. Any representative of this class
can be reduced to the form 
\begin{equation}
f''+(\frac{\gamma}{\xi}+\frac{\delta}{\xi-1}+\frac{\epsilon}{\xi-a})f'+\frac{\alpha\beta\xi-q}{\xi(\xi-1)(\xi-a)}f=0,
\end{equation}
which is known as Heun's differential equation. The coefficients occurring
in this form must satisfy $\alpha+\beta-\gamma-\delta-\epsilon+1=0$;
the constant $q$ is called the accessory parameter. The regular singular
points of this equation lie at $0$, $1$, $a$ and $\infty$. Heun's
differential equation admits $2$ solutions, usually called Heun functions,
which are contained in a set of no less than $192$ different solutions.
One can find a lot of information on these functions, such as for
instance the fact that said functions are often written down as power
series expressions in Riemann's $P$-symbols and thus in series of
hypergeometric functions, which are solutions of Riemann's $P$-differential
equation. The most prominent among these solutions are certainly the
polynomial ones $Hp(a,q;\alpha,\beta,\gamma,\delta;\xi)$, generally
known as Heun polynomials, which are again special types of Heine-Stieltjes
polynomials. The non-polynomial solutions are often denoted by $H\ell(a,q;\alpha,\beta,\gamma,\delta;\xi)$
and rarely written down explicitly, which is mainly due to the fact
that the coefficients have to be determined by solving a three-term
recursion relation. More precisely, the situation is as follows: Since
a Frobenius ansatz of the form $(10)$ leads to the critical exponents
$\rho_{1}=0$ and $\rho_{2}=1-\gamma$, it seems justified to consider
- given the case that $\gamma$ is not a positive integer and there
holds $\vert\xi\vert<1$ - the power series expression $f_{1}(\xi)=H\ell(a,q;\alpha,\beta,\gamma,\delta;\xi)=\overset{\infty}{\underset{k=0}{\sum}}w_{k}\xi^{k}$,
which, after the result $w_{1}:=\ll m_{(1)},m_{(2)}\gg_{1}w_{0}=m_{(1)1}w_{0}=-\frac{q}{a\gamma}w_{0}$
is obtained, leads to the three-term recurrence relation

\begin{equation}
w_{k+2}=m_{(1)k+2}w_{k+1}+m_{(2)k+2}w_{k},
\end{equation}
according to which $m_{(1)k+2}=m_{(1)}(a,q;\alpha,\beta,\gamma,\delta;k+2):=\frac{(k+1)\left\{ (k+\gamma)(a+1)+a\delta+\epsilon\right\} +q}{a(k+2)(k+\gamma+1)}$
and $m_{(2)k+2}=m_{(2)}(a,q;\alpha,\beta,\gamma,\delta;k+2):=-\frac{(k+\alpha)(k+\beta)}{a(k+2)(k+\gamma+1)}$
is assumed to apply by definition. As a direct result, it can be concluded
that the solution of this recurrence relation must be of the form
$(17)$, so that $w_{k+2}=\ll m_{(1)},m_{(2)}\gg_{k+2}w_{0}$; at
least provided that the objects $X_{j}^{a_{j}}=X^{a_{j}}(j)$ have
the components $X_{j}^{0}=\theta(j),$ $X_{j}^{1}=m_{(1)}(j)$ and
$X_{j}^{2}=m_{(2)}(j).$ In particular, the situation is as follows:
For $k=0$, relation $(17)$ reads $\ll m_{(1)},m_{(2)}\gg_{2}\:=W_{a_{1}a_{0}}X_{1}^{a_{1}}X_{0}^{a_{0}}=W_{11}X_{1}^{1}X_{0}^{1}+W_{20}X_{1}^{2}X_{0}^{0}=m_{(1)1}m_{(1)0}+m_{(2)0}$,
which coincides exactly with what is obtained from $(35).$ For $k=2$,
relation $(17)$ reads $\ll m_{(1)},m_{(2)}\gg_{3}\:=W_{a_{2}a_{1}a_{0}}X_{2}^{a_{2}}X_{1}^{a_{1}}X_{0}^{a_{0}}=W_{111}X_{2}^{1}X_{1}^{1}X_{0}^{1}+W_{120}X_{2}^{1}X_{1}^{2}X_{0}^{0}+W_{201}X_{2}^{2}X_{1}^{0}X_{0}^{1}=m_{(1)2}m_{(1)1}m_{(1)0}+m_{(1)2}m_{(2)0}+m_{(2)2}m_{(1)0}$,
which also coincides with what is obtained from $(35).$ By further
iteration, one then easily verifies that solutions of the Heun equation
$(34)$ can actually be written as a power series of the form $(10)$
with coefficients of the form $(16)$ and $(17),$ respectively. Additionally,
one finds that the solution which corresponds to the critical exponent
$1-\gamma$ in the vicinity of the singular point zero reads $f_{2}(\xi)=\xi^{1-\gamma}H\ell(a,(a\delta+\epsilon)(1-\gamma)+q;\alpha+1-\gamma,\beta+1-\gamma,2-\gamma,\delta;\xi).$

Note that Heun's equation and its confluent form as well as their
generalizations \cite{batic2006generalized,schafke2013gewohnliche}
have numerous applications in both mathematics and theoretical physics.
This is not least because Heun functions and their associated confluent
forms cannot only be used for finding solutions to the Mathieu, Lamé,
Whittaker-Hill and Ince equations, they can also be used to determine
solutions of the Schrödinger equation for different types of potentials
in different dimensions \cite{karwowski2014biconfluent,slavjanov2000special}.
Besides that, the said functions can be used to solve both the Klein-Gordon
equation for real and complex scalar fields and the Dirac equation
for both massless and massive Fermions in quantum field theory on
different curved geometric backgrounds as well as the the Regge-Wheeler
equation, the Zerelli equation the and the Teukolsky master equation
\cite{batic2007heun,batic2010hypergeneralized,birkandan2007examples,chandrasekhar1983mathematical,fiziev2006exact,fiziev2010classes,ishkhanyan2016singular,kraniotis2016klein,kraniotis2019massive,misner2017gravitation,regge1957stability,staicova2011spectrum,teukolsky1972rotating,zerilli1970effective,ishkhanyan2015exact}.
For a more detailed overview of the applications of the Heun equation
and its confluent form, see \cite{hortacsu2013heun}.

However, the methods developed in the given work can also be applied
to differential equations with a higher number of regular singular
points, such as, for instance, the generalized Lamé equation, which
has five regular singular points or to the so-called hypergeneralized
Heun equation \cite{batic2010hypergeneralized}, which has six such
points. In fact, it may be concluded that all non-trivial solutions
of equations of the Fuchsian class can be represented in the form
presented in this work. 

Finally, it shall be pointed out that the method for determining the
coefficients of solutions of Fuchsian differential equations presented
in this work also has been successfully applied to another physical
problem, namely the problem of how to calculate the exact structure
of a so-called profile function corresponding to the gravitational
field of a massless ultrarelativistic particle on the event horizon
of a charged rotating black hole \cite{huber2019ultpar}. This problem
leads to a Fuchsian differential equation with five regular singular
points, the so-called generalized Dray-'t Hooft equation, which occurs
as a special case of the generalized Lamé equation. As it turns out,
the coefficients of solutions to this equation are exactly of the
form presented in this work.

\textbf{Comment: }After finishing this paper, it was brought to my attention by an unknown referee
that solutions to the recurrence relations treated in theorems one
and two of the second section of this work had previously been found
by Mallik in \cite{mallik1998solutions}. \newpage{}
\begin{description}
\item [{Acknowledgements:}]~
\end{description}
I want to thank Herbert Balasin for interesting and illuminating discussions
on the subject.

\bibliographystyle{plain}
\addcontentsline{toc}{section}{\refname}\bibliography{0C__Arbeiten_litfuchs}

\end{document}